\documentclass[aps,prb,reprint,twocolumn,superscriptaddress,floatfix,nofootinbib]{revtex4-1}
\usepackage{amsmath,amssymb,color,comment}
\usepackage{graphicx}
\usepackage[english]{babel}
\usepackage[bookmarks=true,colorlinks,linkcolor=blue,urlcolor=blue,citecolor=blue]{hyperref}
\DeclareMathOperator{\Tr}{Tr}
\usepackage{bm}
\usepackage{ulem}
\usepackage{braket}

\newcommand{\rme}{{\rm e}}
\newcommand{\rmi}{{\rm i}}
\newcommand{\Fig}[1]{Fig.~\ref{#1}}

\newcommand{\sz}{\ensuremath{\sigma^{z}}}

\newcommand{\hi}{\ensuremath{h_{\rm i}}}
\newcommand{\hf}{\ensuremath{h_{\rm f}}}
\newcommand{\hce}{\ensuremath{h^{\rm e}_{\rm c}}}
\newcommand{\hczII}{\ensuremath{h^{\rm II}_{\rm c,z}}}
\newcommand{\hcxII}{\ensuremath{h^{\rm II}_{\rm c,x}}}
\newcommand{\hcI}{\ensuremath{h^{\rm I}_{\rm c,z}}}

\def\figpath{.}

\begin{document}

\title{Dynamical phase diagram of quantum spin chains with long-range interactions}

\author{Jad C. Halimeh}
\affiliation{Physics Department and Arnold Sommerfeld Center for Theoretical Physics, Ludwig-Maximilians-Universit\"at M\"unchen, D-80333 M\"unchen, Germany}
\affiliation{Max Planck Institute for the Physics of Complex Systems, 01187 Dresden, Germany}

\author{Valentin Zauner-Stauber}
\affiliation{Vienna Center for Quantum Technology, University of Vienna, Boltzmanngasse 5, 1090 Wien, Austria}


\date{\today}

\begin{abstract}
Using an infinite Matrix Product State (iMPS) technique based on the time-dependent variational principle (TDVP), we study two major types of dynamical phase transitions (DPT) in the one-dimensional transverse-field Ising model (TFIM) with long-range power-law ($\propto1/r^{\alpha}$ with $r$ inter-spin distance) interactions out of equilibrium in the thermodynamic limit -- \textit{DPT-I}: based on an order parameter in a (quasi-)steady state, and \textit{DPT-II}: based on non-analyticities (cusps) in the Loschmidt-echo return rate. We construct the corresponding rich dynamical phase diagram, whilst considering different quench initial conditions. We find a nontrivial connection between both types of DPT based on their critical lines. Moreover, and very interestingly, we detect a new DPT-II dynamical phase in a certain range of interaction exponent $\alpha$, characterized by what we call \textit{anomalous cusps} that are distinct from the \textit{regular cusps} usually associated with DPT-II. Our results provide the characterization of experimentally accessible signatures of the dynamical phases studied in this work.
\end{abstract}
\maketitle

\section{Introduction}
Phase transitions are among the most fascinating phenomena in physics whereby a small change in a control parameter of the system can drive the system between extremely different phases that are not adiabatically connected to one another. This gives rise to non-analyticities in the free energy even when the system itself is described by a completely analytic Hamiltonian without any singularities. Quantum and classical equilibrium phase transitions are textbook subjects that have been very well-studied and established in various systems. Recently, and particularly in the context of closed quantum systems, quench dynamics \cite{Calabrese2011} and post-quench system behavior have received a lot of attention. Of special interest is the concept of DPTs, where, in one type (DPT-I) thereof, critical behavior is inspected in a post-quench (quasi-)steady state, such as a (pre)thermal state at (intermediate) long times. Reaching a (quasi-)steady state is crucial in DPT-I in order to extract a steady-state value of the order parameter under consideration. On the other hand, a second type (DPT-II) was defined in the seminal work Ref.~\onlinecite{Heyl2013}, and has been studied extensively analytically \cite{Pozsgay2013,Heyl2014,Hickey2014,Vajna2014,Vajna2015,Heyl2015,Schmitt2015,Campbell2016,Heyl2016,Zunkovic2016} and numerically \cite{Karrasch2013,Fagotti2013,Canovi2014,Kriel2014,Andraschko2014,Sharma2015,Zhang2016,Sharma2016,Vid2016,Homrig17} 
in several models, and has just recently even been observed experimentally. \cite{Peng2015,Flaeschner2016} DPT-II also involves a quench between an initial and a final Hamiltonian, however, unlike DPT-I, reaching a steady state is not a requirement, not least because DPT-II actually manifests itself as non-analyticities in the Loschmidt-echo return rate \cite{Heyl2013} as a function of evolution time, and does not in principle rely on a Landau-type order parameter. In general, DPT-II occurs when an initial state undergoes a quench where the control parameter crosses the quantum equilibrium critical point,\cite{Heyl2013,Heyl2014,Heyl2015} and has been observed in the nearest-neighbor TFIM (NN-TFIM) with such quenches from both equilibrium phases.\cite{Heyl2013,Heyl2015} However, there are exceptions to this rule, where DPT-II occurs for quenches within the same equilibrium phase and is absent in quenches across the quantum equilibrium critical point. \cite{Andraschko2014,Vajna2014}

\begin{figure}[t]
 \centering
 \includegraphics[width=\linewidth]{\figpath/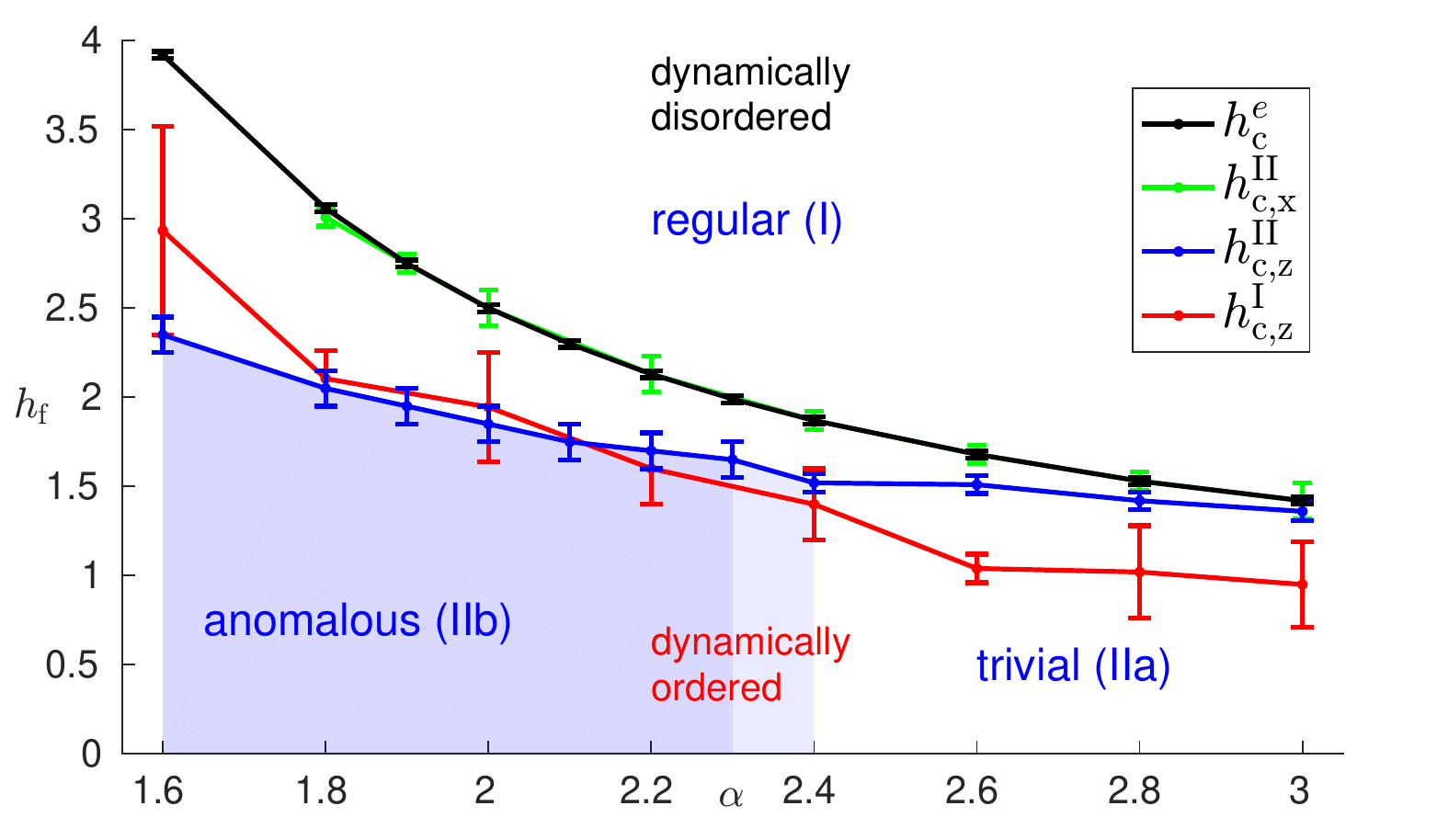}
 \caption{Dynamical phase diagram of the LR-TFIM: $h^{\rm e}_{\rm c}$ is the equilibrium critical line, $h^{\rm I}_{\rm c,z}$ is the DPT-I critical line, $h^{\rm II}_{\rm c,z(x)}$ is the DPT-II critical line for quenches from $h_{\rm i}=0$ ($h_{\rm i}\to\infty$) which signifies the onset of \textit{regular cusps} for quenches above (below) it. Note how the critical lines $h^{\rm II}_{\rm c,x}$ and $h^{\rm e}_{\rm c}$ overlap very well within the precision of our numerical simulations. For quenches from $h_{\text{i}}=0$ to below $h_{\text{c,z}}^{\text{II}}$, the system exhibits a trivial (cusp-free) phase for $\alpha\gtrsim2.3$ and an anomalous phase for $\alpha\lesssim2.3$. The dynamically ordered and disordered phases are related to DPT-I \cite{Halimeh2016} and are separated by $\hcI$. For a discussion of the error bounds see Sec.~\ref{sec:ErrorBounds}.
 }
 \label{fig:P1_PhaseDiag}
\end{figure}

In a related study \cite{Halimeh2016} we demonstrate that DPT-I in the long-range TFIM (LR-TFIM) exists in situations even when the system under consideration does not exhibit a finite-temperature phase transition.\cite{Dutta2001} For further details, we refer the reader to Ref.~\onlinecite{Halimeh2016}. In this work, we shall focus on the behavior of DPT-II, its detection in the framework of iMPS \cite{MPS1,MPS2,Stauber2017} in the thermodynamic limit, the characterization of its different phases, and its relation to DPT-I. Our results are summarized in the dynamical phase diagram shown in \Fig{fig:P1_PhaseDiag}. We find it very advantageous to work with iMPS here as opposed to finite-size time-dependent density matrix renormalization group ($t$-DMRG) methods \cite{TEBD,tDMRG1,tDMRG2} in order to see actual non-analytic cusps in the Loschmidt-echo return rate,\cite{Heyl2013} which in finite systems are inherently nonexistent and thus have to be extracted through a finite-size scaling procedure (for technical details, see Ref.~\onlinecite{supp}). 

The paper is organized as follows: In Sec.~\ref{sec:model} we present the main model used in the iMPS simulations for this work, and additionally review the Loschmidt-echo return rate, the motivation behind it, and the types of quench that give rise to it. In Sec.~\ref{sec:results} we present the main results of this paper, and introduce and characterize the \textit{anomalous dynamical phase}. We conclude in Sec.~\ref{sec:conclusion}

\begin{figure}[t]
 \centering
 \includegraphics[width=\linewidth]{\figpath/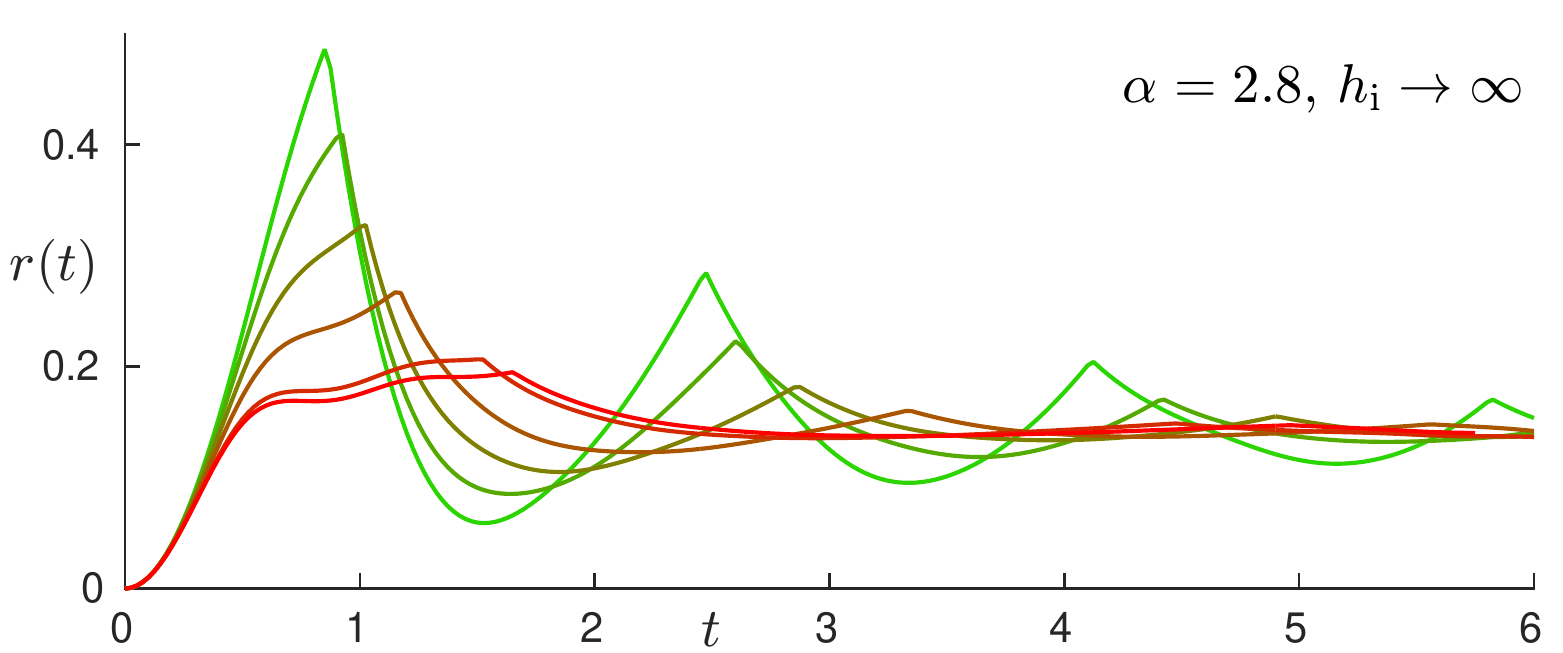}
 \caption{Return-rate function $r(t)$ for $\alpha=2.8$, quenches from $h_{\text{i}}\to\infty$ to $h_{\text{f}}\in [0.75, 1.40]$ (from green to red).
 }
 \label{fig:P2_xup_cusps}
\end{figure}

\section{Model and return rate}
\label{sec:model}
We consider the LR-TFIM with power-law interactions \cite{LRTFI1,LRTFI2,LRTFI5,LRTFI3,LRTFI4}

\begin{equation}\label{eq:LRTFIM}
\mathcal{H}=-J\sum_{j>i=1}^{L}\frac{\sigma^z_i\sigma^z_j}{|i-j|^{\alpha}}-h\sum_{i=1}^{L}\sigma^x_i,
\end{equation}

\noindent where $\sigma^{x,z}_{i}$ are Pauli matrices acting on site $i$, $J$ is the spin-spin coupling constant, $h$ is the magnetic field, $L$ is the number of sites, and we consider the ferromagnetic case $J>0$ in the thermodynamic limit $L\to\infty$. The efficient implementation of the long-range interactions is based on Ref.~\onlinecite{Crosswhite2008}.

For quenches whose time evolution is propagated by $\mathcal{H}$, the motivation \cite{Heyl2013} for studying non-analyticities in the Loschmidt amplitude,\cite{Silva2008} \textit{i.e.}~the overlap between the initial and time-evolved states

\begin{equation}
G(t)=\braket{\psi(0)|\psi(t)}=\braket{\psi(0)|e^{-\rmi \mathcal{H} t}|\psi(0)},
\label{eq:L_echo}
\end{equation}
is to exploit the similarity between \eqref{eq:L_echo} and the partition function $\mathcal{Z}(\beta)=\Tr \rme^{-\beta\mathcal{H}}$ of the system in thermal equilibrium 
at inverse temperature $\beta$, and interpret \eqref{eq:L_echo} as a boundary partition function with boundary conditions $\ket{\psi(0)}$ and complex inverse temperature $z$
\begin{equation}
\mathcal{Z}_{b}(z)=\braket{\psi(0)|e^{-z \mathcal{H}}|\psi(0)},
\label{eq:bdry_Z}
\end{equation}
along the imaginary axis $z=\rmi t$. 
With the return probability (Loschmidt echo) $\mathcal{L}(t)=|G(t)|^{2}$, the \textit{return rate function}
\begin{equation}
r(t)=-\lim_{L\to\infty}\frac{1}{L}\ln|G(t)|^2,
\label{eq:ratefun}
\end{equation}
can thus be construed as an analogue of the free energy per site, in which non-analyticities indicate the presence of DPT-II, thereby making a connection between finite-temperature partition functions and time evolution, asking whether the latter can exhibit phase transitions as well.\cite{footnote} DPT-II has been studied in various systems, and in the case of the transverse-field Ising model (TFIM), it has been shown for the integrable cases of nearest-neighbor \cite{Heyl2013,Heyl2015} ($\alpha\to\infty$) and infinite-range interactions ($\alpha=0$).\cite{Zunkovic2016,Homrig17} In this work, we study DPT-II in the one-dimensional nonintegrable ferromagnetic LR-TFIM for $1<\alpha<\infty$ in the thermodynamic limit.

\begin{figure}[t]
 \centering
 \includegraphics[width=\linewidth]{\figpath/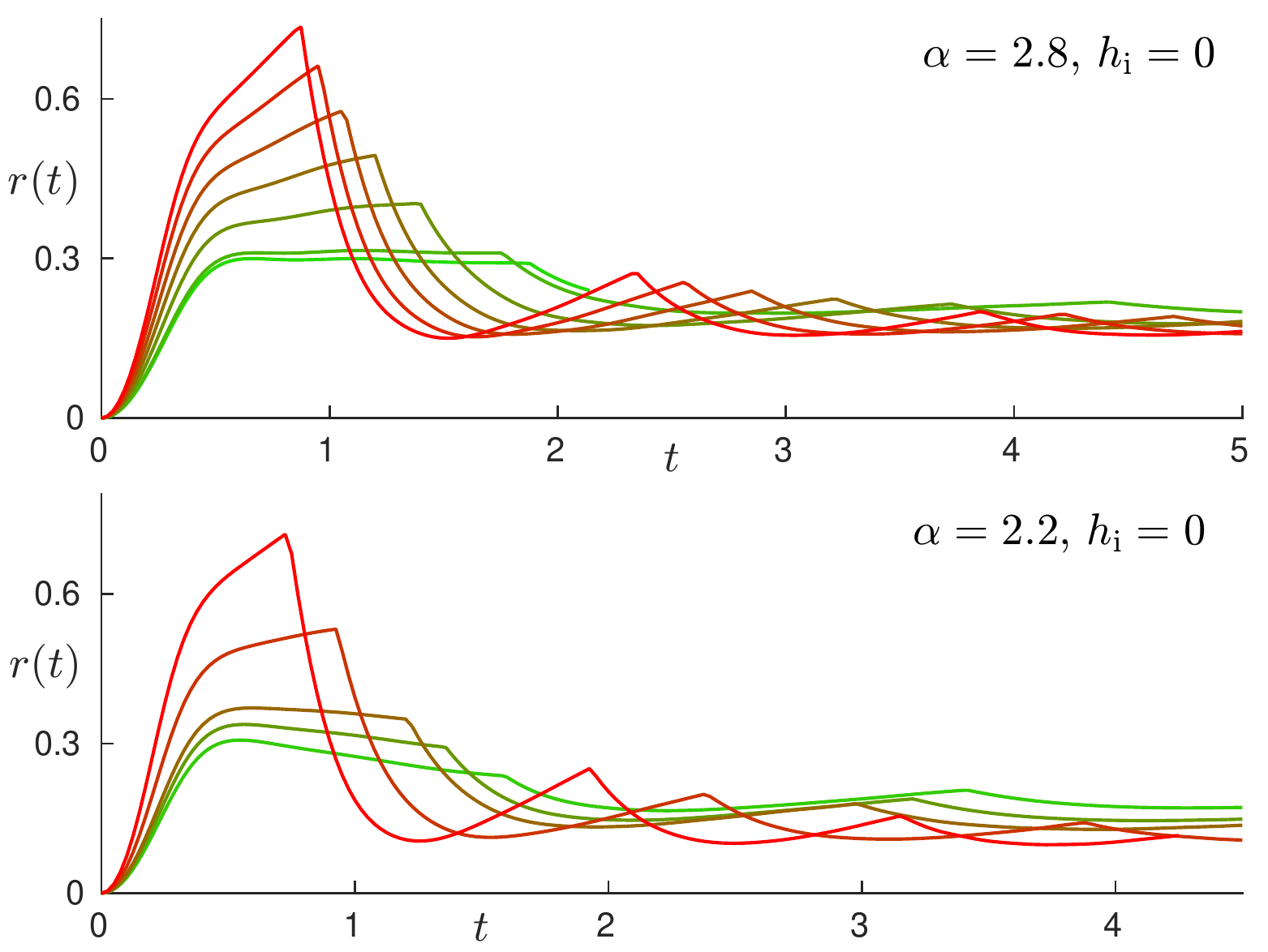}
 \caption{Return-rate function $r(t)$ for quenches from $\hi=0$ to $\hf\in [1.50, 2.30]$ (from green to red) for $\alpha=2.8$ (top) and $\hf\in [1.80, 2.80]$ (from green to red) for $\alpha=2.2$ (bottom). This behavior is qualitatively the same for all $\alpha$.
 }
 \label{fig:P3_zup_reg}
\end{figure}

To study DPT-II, we calculate the return rate function per site \eqref{eq:ratefun}, after performing a quantum quench, where we prepare the system in the groundstate $|\psi(0)\rangle$ of $\mathcal{H}(h_{\text{i}})$ (that is~\eqref{eq:LRTFIM} with $h=h_{\text{i}}$), and then abruptly change the magnetic field from $h_{\text{i}}$ to $h_{\text{f}}\neq h_{\text{i}}$ at time $t=0$. As the system evolves in time as propagated by  $\mathcal{H}(h_{\text{f}})$, the return rate function per site~\eqref{eq:ratefun} can then be calculated from the overlap~\eqref{eq:L_echo} of the initial state with its time-evolved self. The return rate function can be calculated efficiently directly in the thermodynamic limit with iMPS techniques as (minus) the logarithm of the dominant eigenvalue of the (mixed) Matrix Product State (MPS) transfer matrix 
arising in the overlap between the initial state and the time-evolved state at time $t$. Cusps in the return rate occur when there is a level crossing of the dominant eigenvalues of this transfer matrix. For the technical details of this method, we refer the reader to Ref.~\onlinecite{supp}.

\section{Results and discussion}
\label{sec:results}
We shall now discuss the results of our numerical simulations for two types of quenches in the LR-TFIM \eqref{eq:LRTFIM} and extract signatures of criticality for DPT-II: (i) A quench from $h_{\rm i}\to\infty$, corresponding to an initial state that is completely polarized in the positive $x$-direction (X-quench), and (ii) a quench from $h_{\rm i}=0$, where the groundstate is two-fold degenerate. We choose the state completely polarized in the positive $z$-direction as initial state (Z-quench).
 
 For the X-quenches we find a critical phase with the occurrence of conventional cusps in the return rate \eqref{eq:ratefun}, as first observed in the NN-TFIM.\cite{Heyl2013} Henceforth, these cusps will be called \textit{regular} cusps. We also find a trivial phase with no cusps in the return rate. These two phases exist for all studied $\alpha$.  For Z-quenches we again find a regular and a trivial phase, but additionally encounter an anomalous phase, replacing the trivial phase for $\alpha\lesssim2.3$ only. The \textit{anomalous} cusps appearing in this phase are qualitatively different from the regular ones.  Lastly, signatures of criticality for DPT-I, studied and characterized in Ref.~\onlinecite{Halimeh2016}, are also included for comparison. See \Fig{fig:P1_PhaseDiag} for the dynamical phase diagram.

First, let us consider the case of X-quenches to some final value $\hf$ of the transverse field, where we encounter the same situation as for the NN-TFIM.\cite{Heyl2013} For quenches within the disordered \textit{equilibrium} phase $\hf>\hce(\alpha)$, with $\hce(\alpha)$ the quantum equilibrium critical point \cite{footnote0} at the given $\alpha$, we observe a trivial phase with no cusps at all in the return rate. However, quenching across $\hce(\alpha)$ into the ordered equilibrium phase, we encounter a regular critical phase and (regular) cusps appear. \Fig{fig:P2_xup_cusps} shows an example for $\alpha=2.8$ and various $\hf$ across the equilibrium critical point. It is apparent that the deeper the quench into the ordered phase, the more pronounced the cusps are, and the smaller the time intervals between the cusps become. As the quench approaches the critical point from below, these cusps appear less sharp and the intervals between them get longer. All cusps completely disappear simultaneously when crossing $\hce(\alpha)$ from below, i.e. when quenching within the disordered phase. The similarity to the NN-TFIM case is indicated in \Fig{fig:P1_PhaseDiag} by the overlap of the DPT-II critical line $\hcxII$ for X-quenches with the quantum equilibrium critical line $\hce$ within the precision of our numerical simulations. 

\begin{figure}[t]
 \centering
 \includegraphics[width=\linewidth]{\figpath/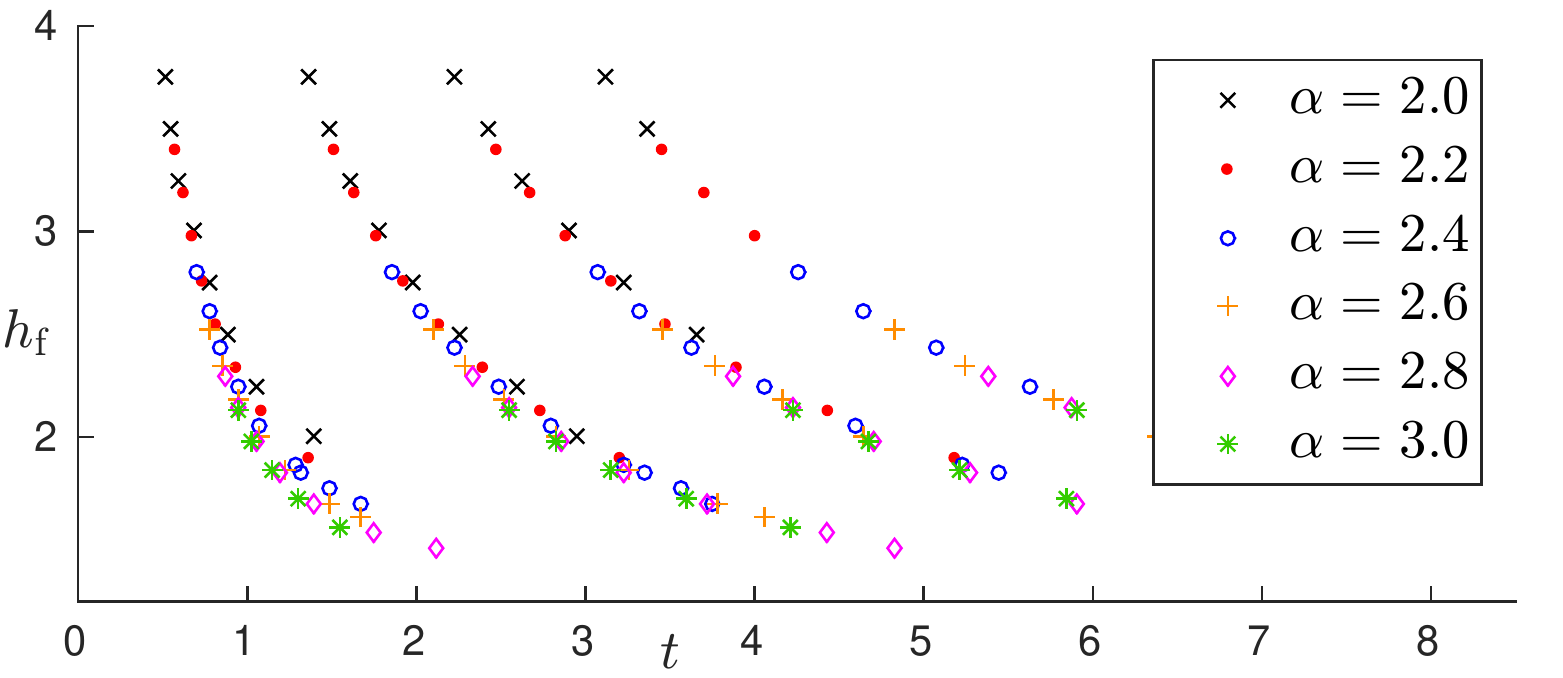}
 \caption{Times of regular cusps (c.f. \Fig{fig:P3_zup_reg}), appearing in quenches from $\hi=0$, for various $\alpha\in[2,3]$ and $\hf>\hczII$. It is clearly visible that the respective times of the cusps \textit{decrease} with increasing $\hf$. Moreover, the dependence of the times for each cusp seems independent of $\alpha$, within an uncertainty of $\approx5\%$. Missing points at higher times, especially for low $\hf$, are due to limited simulation times.
 }
 \label{fig:P4_zup_reg_t}
\end{figure}

On the other hand, for Z-quenches, we see three distinct DPT-II phases: (I) a \textit{regular phase} with only regular cusps in the return rate (as for X-quenches above), that occurs when quenching across a DPT-II critical field value $\hczII(\alpha)$, which is however smaller than $\hce(\alpha)$ (c.f. \Fig{fig:P1_PhaseDiag}). For quenches below $\hczII(\alpha)$ we encounter (IIa) a \textit{trivial phase} for $\alpha\gtrsim2.3$, that exhibits no cusps at all in the return rate, and interestingly, (IIb) an \textit{anomalous} phase for $\alpha\lesssim2.3$, that exhibits anomalous cusps in the return rate that are qualitatively different from the regular cusps in phase (I). The additional appearance of phase (IIb) featuring anomalous cusps and a critical field $\hczII(\alpha)$ different from $\hce(\alpha)$ for this quench are the two major differences to such a quench in the NN-TFIM.

\begin{figure}[t]
 \centering
 \includegraphics[width=\linewidth]{\figpath/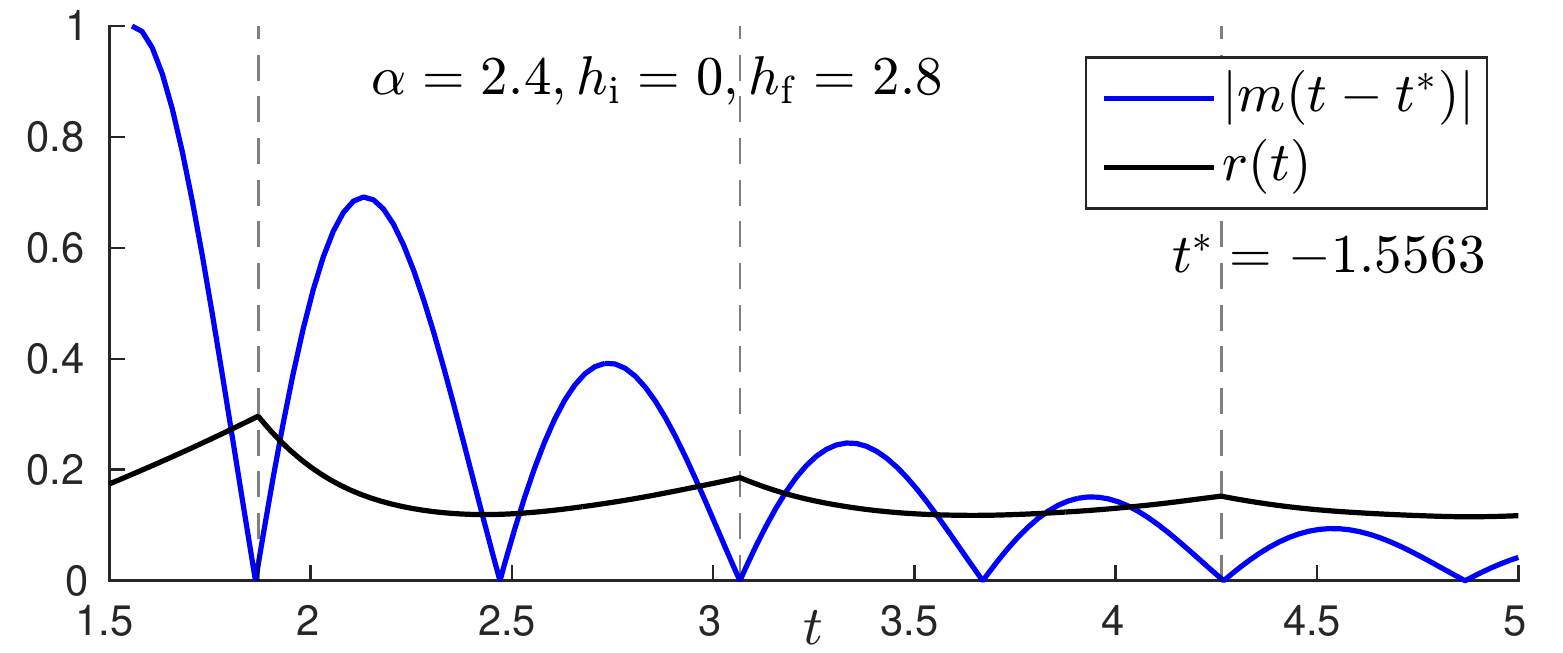}
 \caption{Return rate function $r(t)$ plotted together with the (absolute value of the) order parameter $|m(t)|$. The cusps in the return rate coincide with zeros of the shifted order parameter. 
 }
 \label{fig:P5_ZvsRF}
\end{figure}

Let us focus on the regular phase (I) first. In \Fig{fig:P3_zup_reg} we show results for Z-quenches for various $\alpha$ and $h_{\rm f}>\hczII$, where cusps in the rate function are separated by roughly equal time intervals. With lowering $\hf$ and approaching $\hczII$, these time intervals \textit{increase} and the cusps get less sharp until they all vanish simultaneously upon crossing $\hczII$. The time intervals as a function of $\hf$ seem to be largely independent of $\alpha$ (see \Fig{fig:P4_zup_reg_t}) and are also reflected in the oscillation period of the order parameter $m(t)=\braket{\sz(t)}$. The times at which regular cusps appear match the zero crossings of $m(t)$ up to a temporal shift, as shown in \Fig{fig:P5_ZvsRF}. This fact has already been observed in the NN-TFIM \cite{Heyl2013,Karrasch2013} and next-nearest-neighbor TFIM.\cite{Karrasch2013} It is worth noting here that the periodicity of the return rate is doubled \cite{Karrasch2013, Heyl2013} if one considers the return rate with respect to the degenerate subspace of the initial state rather than the initial state itself as we do. Thus, the only difference of this phase to conventional DPT-II criticality in the NN-TFIM is the critical field $\hczII$. 

\begin{figure}[t]
 \centering
 \includegraphics[width=\linewidth]{\figpath/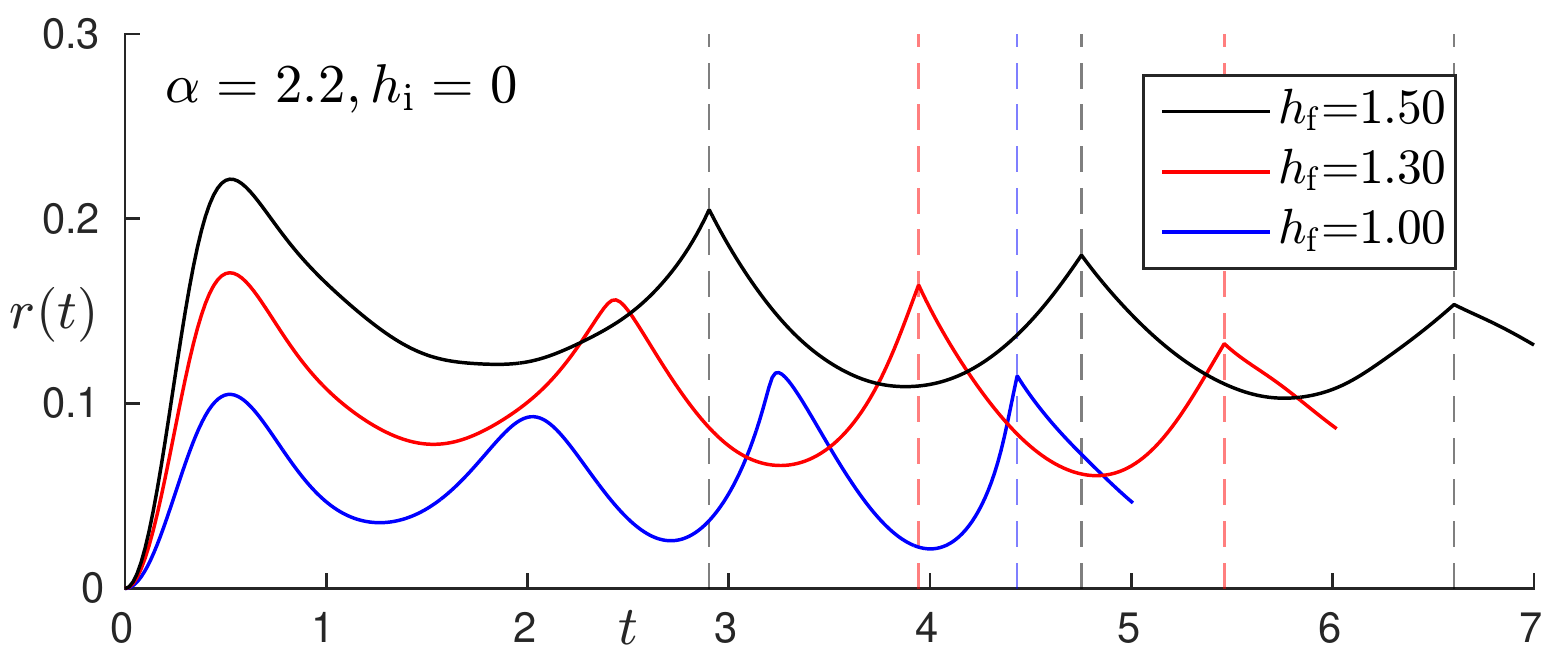}
 \caption{Examples of anomalous cusps (marked by vertical dashed lines) for $\hf<h^{\rm II}_{\rm c,z}$. It is apparent that with increasing $h_{\rm f}$ more such cusps develop at smaller times, while their respective locations however move to higher times (c.f. also \Fig{fig:P7_anom_t}).
 }
 \label{fig:P6_zup_anom}
\end{figure}

\begin{figure}[t]
 \centering
 \includegraphics[width=\linewidth,keepaspectratio=true]{\figpath/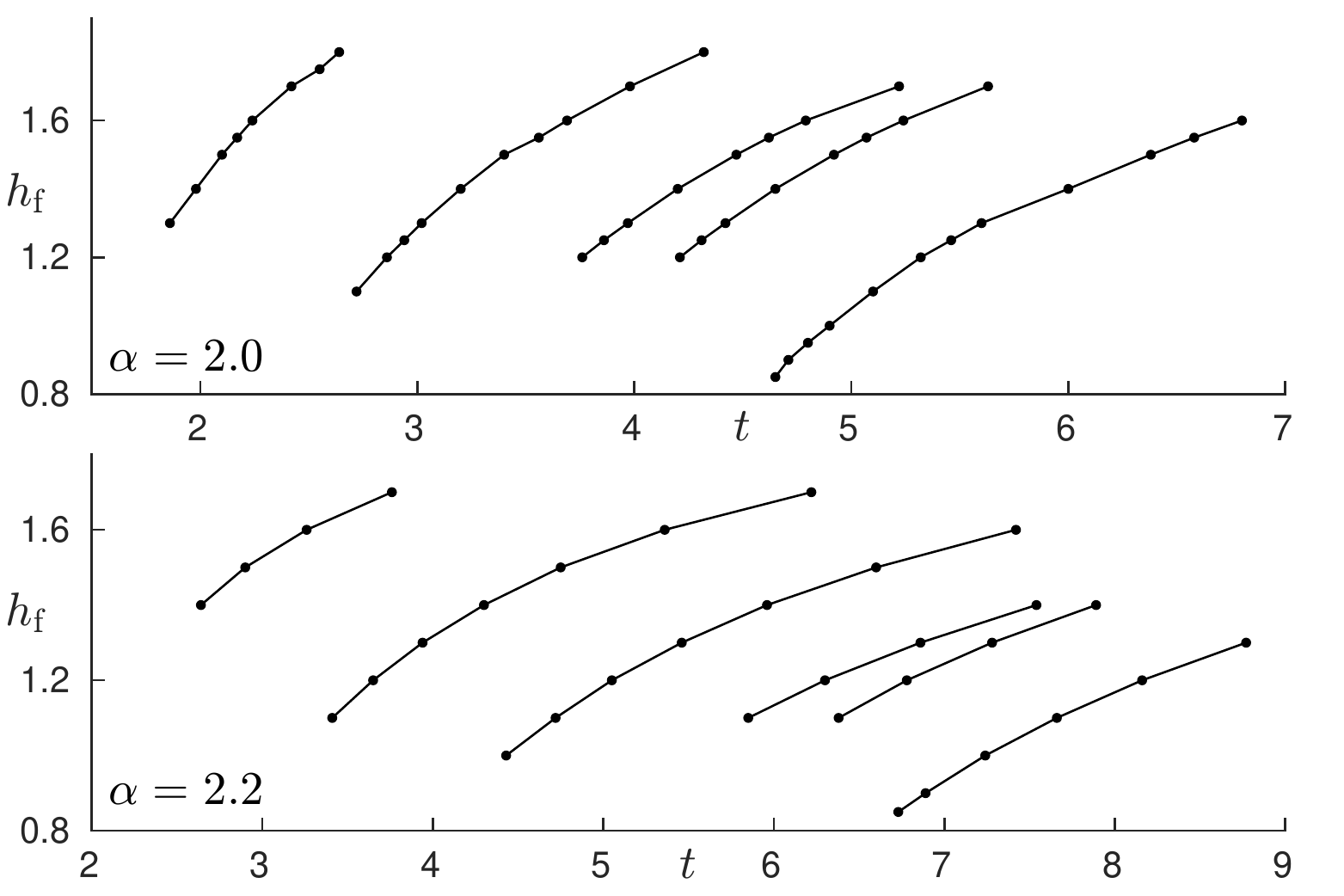}
 \caption{Times of anomalous cusps for $\alpha=2$ (top) and $\alpha=2.2$ (bottom) as a function of $h_{\rm f}<h^{\rm II}_{c,z}$. For $h_{\rm f}>h^{\rm II}_{c,z}$, the anomalous cusps vanish as the regular cusps take over. Missing points at higher times are due to limited simulation times, the terminated lines would continue further to the top right.
 }
 \label{fig:P7_anom_t}
\end{figure}

The regular critical phase goes over into an anomalous critical phase (IIb) for quenches below $\hczII$ and $\alpha\lesssim 2.3$. There, after a coexistence region with the regular cusps in a finite range of values of $h_{\rm f}$ around $\hczII$, a qualitatively different type of cusps in the rate function appears (see \Fig{fig:P6_zup_anom}). Upon further lowering $\hf$, the time intervals between these cusps \textit{decreases}, contrary to the regular cusps in phase (I), and they vanish one by one as $\hf\to 0$, starting at early times, as can be seen in \Fig{fig:P6_zup_anom} and \Fig{fig:P7_anom_t}. As our evolution times for accurate simulations are limited, we can only conjecture here that this type of cusps exists for any small $\hf>0$, albeit only appearing at very large times. This has in fact been confirmed in the $\alpha\to0$ limit in a follow-up paper to this work.\cite{Homrig17} It is worth mentioning that some of these anomalous cusps show a ``double-cusp'' structure, where the location of these double cusps also drifts with $\alpha$. This behavior is showcased in Ref.~\onlinecite{supp} and was also observed in Ref. \onlinecite{Karrasch2013}.

We emphasize here that the two types of observed cusps are qualitatively different, as can be seen in how they arise from two qualitatively different groups of rate-function branches calculated from the MPS transfer matrix where it is always the lowest branch that corresponds to the actual return rate function.\cite{supp} 
The characteristic feature distinguishing the anomalous phase from the regular phase is that its cusps only appear \textit{after} the first minimum in the return rate, with more and more smooth peaks preceding the first cusp for smaller and smaller quenches. Regular cusps on the other hand always appear over the entire time range, with the first cusp always appearing before the first minimum. Additionally, the time intervals between the crossings behave differently in both phases upon varying $\hf$. Ref.~\onlinecite{supp} provides further technical details on the crossover between the regular phase (I) and the anomalous phase (IIb), and discusses the physical origin of these two different types of cusps to different groups of Fisher-zero lines of \eqref{eq:bdry_Z} in the complex plane \cite{Heyl2013} cutting the imaginary axis in different ways.

Finally, we discuss the relationship between DPT-I and DPT-II according to their critical lines in the rich dynamical phase diagram of \Fig{fig:P1_PhaseDiag}. The critical line of DPT-I is much harder to obtain, as this kind of DPT relies on reaching a (quasi-)steady state at time $t\approx\tau$, from which the order parameter $\tilde{m}=m(\tau)$ is extracted, and then one tries to establish the existence or absence of a non-analyticity of this order parameter as a function of $\hf$, as in our case. DPT-I in the nonintegrable LR-TFIM has been extensively studied,\cite{Halimeh2016} and it was determined that prethermalization conspires to give rise to DPT-I even for $\alpha>2$ where the LR-TFIM exhibits no thermal phase transition in one spatial dimension.\cite{Dutta2001} Comparing the DPT-I and DPT-II critical lines, an unequivocal conclusion is unrealistic, given the evolution times of accurate simulations reached, or by those also carried out using finite-size $t$-DMRG.\cite{Halimeh2016} However, from the data, it seems that the two types of DPT are nontrivially connected and at least show the same tendency in their $\alpha$-dependence.

\section{Conclusion}
\label{sec:conclusion}
In summary, we have carried out time-evolution simulations of pure quantum states after global quantum quenches in the one-dimensional LR-TFIM, and studied the corresponding post-quench DPT-I and DPT-II phases in the thermodynamic limit using an iMPS technique based on TDVP.  Within the precision of our numerics, we find that the DPT-II critical line for X-quenches overlaps with the quantum equilibrium critical line separating two phases, one displaying (regular) cusps in~\eqref{eq:ratefun} when quenching across this line, and a phase with no cusps otherwise. For Z-quenches, the critical lines for both DPT-I and DPT-II seem to show a nontrivial connection. Whereas in DPT-I two dynamical phases (ordered and disordered) appear for all $\alpha$, we find three distinct DPT-II phases: a \textit{regular} phase (I) for quenches above the line for all $\alpha$ where regular cusps appear in~\eqref{eq:ratefun}, a trivial phase (IIa) for quenches below the line for $\alpha\gtrsim2.3$ where no cusps appear, and a new \textit{anomalous} phase (IIb) for quenches below the line for $\alpha\lesssim2.3$ characterized by \textit{anomalous cusps} that are qualitatively different from their regular counterparts present in (I). Finally, Ref.~\onlinecite{LRTFI3} reports on two distinct types of equilibrium universality in the LR-TFIM for $\alpha<2.25$ and $\alpha>2.25$. It would be interesting to know if the appearance of the anomalous DPT-II phase for $\alpha\lesssim2.3$ is connected to this change in universality. We leave this open for future study.

While completing this manuscript, we became aware of a study \cite{Zunkovic2016b} that discusses some of our results in part on finite-size systems. 

\section*{Acknowledgments}
Both authors contributed equally to this work. We thank Damian Draxler, Jutho Haegeman, Michael Kastner, Andreas L\"auchli, Ian P. McCulloch, Francesco Piazza, and Frank Verstraete for inspiring and helpful discussions. V.~Z.-S. gratefully acknowledges support from the Austrian Science Fund (FWF): F4104 SFB
ViCoM and F4014 SFB FoQuS. The computational results presented have been achieved in part using the Vienna Scientific Cluster (VSC).

\appendix

\section{Error bounds for Fig.~\ref{fig:P1_PhaseDiag}}\label{sec:ErrorBounds}
In this section we discuss the error bounds of the dynamical phase diagram displayed in Figure 1 of the main text. $h_{\rm c}^{\rm e}$ has been determined by performing iMPS ground state calculations with varying $h$ while monitoring the expectation value of the order parameter. $h_{\rm c}^{\rm e}$ is then determined as the largest $h$ for which the order parameter is non-zero, and the error is given by the step size in $h$. $h_{\rm c}^{\rm I}$ has been determined by extrapolating the order parameter $m(t)=\braket{\sigma_{i}^{z}(t)}$ to the long-time limit using the fit procedure presented in Ref.~\onlinecite{Halimeh2016} and $h_{\rm c}^{\rm I}$ is again the largest $h_{\rm f}$ for which the extrapolated order parameter is non-zero. Due to limited simulation times the extrapolated values have large statistical error bars, resulting in error bars for $h_{\rm c}^{\rm I}$ which are considerably larger than the step size of $h_{\rm f}$ (see also Figures 3 and 4 in Ref. \onlinecite{Halimeh2016}). Tighter error bars are only achievable with tremendously increased numerical efforts. However, recently a promising new approach \cite{Leviatan2017_supp} could permit additional qualitative statements about the behavior of the order parameter at large times.
$h^{\rm II}_{\rm c,x}$ and $h^{\rm II}_{\rm c,z}$ have been determined by monitoring the appearance of cusps in the the return rate function $r(t)$ \eqref{eq:ratefun}. They are defined as the largest $h_{\rm f}$ for which $r(t)$ shows no cusps (for $h^{\rm II}_{\rm c,z}$ only in the case $\alpha>2.3$). The error bars are thus given by the step size of $h_{\rm f}$.
In the case $\alpha<2.3$, $h^{\rm II}_{\rm c,z}$ is defined as the $h_{\rm f}$ for which anomalous cusps in $r(t)$ are superseded by regular cusps. As this process takes place within a finite range of $h_{\rm f}$ values (see also Ref. \onlinecite{supp}), $h^{\rm II}_{\rm c,z}$ is taken at the center with error bars estimating this regime.


\end{document}